\documentclass[11pt]{article}
\usepackage{a4wide}
\pdfoutput=1

\usepackage{framed}

\usepackage{Packages}

\allowdisplaybreaks[1]

\usepackage{Definitions}

\usepackage{comment}
\usepackage{setspace}

\newcommand{\OfficialTitle}{Warped~anti-de~Sitter~spaces from brane intersections in type II string theory}

\hypersetup{pdfauthor={Domenico Orlando, Linda I. Uruchurtu},pdftitle={\OfficialTitle}}

\author{
  \begin{minipage}{.97\linewidth}
    \vspace{1cm}
    \begin{center}
      \begin{small}
        \textbf{Domenico Orlando}${}^{\#}$ and \textbf{Linda I. Uruchurtu}${}^{\dagger}$
      \end{small}
    \end{center}
    \vspace{1cm}
    \begin{center}
      \begin{minipage}{.7\linewidth}
        {\it \begin{footnotesize}
            \begin{itemize}
            \item[${}^{\#}$]Institute for the Mathematics and Physics of
              the Universe, \\The University of Tokyo, Kashiwa-no-Ha
              5-1-5, \\ Kashiwa-shi, 277-8568 Chiba, Japan. \\
            \item[${}^{\dagger}$] Theoretical Physics Group, The Blackett Laboratory\\
            Imperial College London, Prince Consort Road \\
            London, SW7 2AZ, UK.
            \end{itemize}
          \end{footnotesize}}
      \end{minipage}
    \end{center}
    \vspace{1cm}
  \end{minipage}
}

\date{}

\title{\vspace{2cm}
  \begin{huge}
    \textbf{\OfficialTitle}
  \end{huge}
}

\begin{document}

\numberwithin{equation}{section}

\begin{titlepage}
  \maketitle
  \thispagestyle{empty}

  \vspace{-14cm}
  \begin{flushright}
   IPMU10-0041\\
   Imperial/TP/2010/LIU/01
  \end{flushright}

  \vspace{14cm}

  \begin{center}
    \textsc{Abstract}
  \end{center}
    
  We consider explicit type II string constructions of backgrounds
  containing warped and squashed anti--de Sitter spaces. These are
  obtained via T--duality from brane intersections including
  dyonic black strings, plane waves and monopoles. We also study the
  supersymmetry of these solutions and discuss special values of the
  deformation parameters.

 \end{titlepage}

\setstretch{1.1}


\section{Introduction}

The study of three--dimensional maximally symmetric spaces has
attracted a lot of attention over the years. They arise as near
horizon geometries of various D--brane configurations and in some cases
they can be shown to be exact string backgrounds. Furthermore,
anti--de~Sitter spaces have been used in the context of the AdS/CFT
correspondence~\cite{Maldacena:1998re} leading to a dramatic progress
in the understanding of gauge theories, and have found applications
spanning from black hole physics to condensed matter physics or
strongly coupled plasmas.

Recently, the subject has met with renewed
interest~\cite{Nutku:1993eb,Compere:2008cv} in the context of the
study of Topologically Massive Gravity (TMG) with a negative
cosmological constant~\cite{Deser:1981wh,Deser:1982vy,Chow:2009km}. 
The theory admits a family of asymptotically $\AdS$ solutions
parameterized by the value of the Chern--Simons coupling. Moreover, it was
found to contain two stable warped $\AdS$ vacuum solutions for every
value of the Chern--Simons coupling~\cite{Anninos:2009zi}, where the
term warped stands for the fact that $\AdS_{3}$ is viewed as a Hopf
fibration over $\AdS_{2}$, with the fiber being lengthened by a
constant factor\footnote{For clarifications on the use of terminology with respect to the use of the terms ``squashed'' and ``warped'' we direct the
reader to \cite{Chow:2009km}. }.

In this context a \emph{warped $\AdS_3$} geometry is obtained by
changing the radius of the $S^1$ fiber over $\AdS_2$. In a suitable
coordinate system, we are considering Minkowskian three--manifolds
endowed with a one--parameter family of metrics
\begin{equation}
  \di s^2 [\WAdS_3] = R^2 \left[ \di \omega^2 - \cosh^2 \omega \di \tau^2 + \frac{1}{\cosh^2 \Theta_w} \left( \di \beta + \sinh \omega \di \tau \right)^2 \right] \, ,  
\end{equation}
where $\Omega_w$ is the deformation parameter that interpolates
between $\AdS_3$ for $\Theta_w = 0$ and $\AdS_2 \times S^1$ for
$\Theta_w \to \infty$. Similarly, one can consider a \emph{squashed sphere} obtained by changing the radius of the $S^1 $ fiber over $S^2$. This is described by the one--parameter family of Euclidean metrics
\begin{equation}
  \di s^2[\SqS^3] = R^2 \left[ \di \theta^2 + \sin^2 \theta \di \phi^2 + \frac{1}{\cosh^2 \Theta_m} \left( \di \alpha + \cos \alpha \di \phi \right)^2 \right] \, ,  
\end{equation}
where $\Theta_m$ interpolates between $S^3$ (for $\Theta_m = 0$) and
$S^2 \times S^1$ (for $\Theta_m \to \infty$).

Squashed spacetime geometries are not new and have been studied in the
context of deformed CFTs which were partially motivated by the search for black holes
that generalised BTZ--type backgrounds. A string theory realisation
of metrics including three--spheres and warped $\AdS_3$ spaces was
presented
in~\cite{Israel:2003cx,Israel:2004vv,Detournay:2005fz}. There, such
configurations were obtained as exact marginal deformations of $SU(2)$
and $SL(2,\setR)$ Wess--Zumino--Witten models, thus providing by
construction a worldsheet theory. In particular, it has been shown how
to compute the partition function in the compact case and the spectrum
of the primary operators in the non--compact case. Such
configurations, that rely on a non--vanishing \textsc{ns--ns} field, are not
the subject of this note but we expect them to be related by
S--duality to the ones that we will describe.

It is natural  to ask if one can use the AdS/CFT
correspondence to study string theory on these three-dimensional
warped/squashed backgrounds. Initial studies where made in the context
of TMG in~\cite{Anninos:2008fx}, since the thermodynamic properties
pointed at the existence of a two dimensional boundary CFT dual to the
geometry. Further examples were analysed in~\cite{Anninos:2008qb} as
to include warped solutions arising from theories of Einstein gravity
(such as Topologically Massive Electrodynamics and gravity coupled to
a perfect fluid) and string theory~\cite{Detournay:2005fz,
  Compere:2008cw}. Particular attention has been given to G\"{o}del
black holes, as their non--flat part can be interpreted as resulting
from the squashing of AdS lightcones~\cite{Rooman:1998xf} and are
known to represent exact string theory backgrounds
\cite{Israel:2003ry}.

In all of the examples above it was possible to define an asymptotic
symmetry algebra leading to well--defined conserved charges. The value
of the central charge was then compared to that obtained from the
Bekenstein--Hawking entropy. These observations and the fact that black
strings with near--horizon geometries $\WAdS_{3}\times S^{3}$ arise as
Hopf--T dualizations~\cite{Duff:1998cr} of dyonic black strings in six
dimensions (which in turn has an $\AdS_{3}\times S^{3}$ horizon) lead
to the following duality chain \cite{Anninos:2008qb}: 
\begin{equation}
\label{dualitychainhopf}
  CFT_{1} \longleftrightarrow \AdS_{3}\times S^{3}_{1} \buildrel \text{Hopf}  \over \longleftrightarrow \WAdS_{3}\times S^{3}_{2} \longleftrightarrow CFT_{2} \, .
\end{equation} 
However, it was pointed out in the same paper, that there was no
D--brane interpretation of the Hopf T--dual black string with near
horizon geometry $\WAdS_{3}\times S^{3}$. A step towards that
direction was given in~\cite{Levi:2009az}, where G\"{o}del space
emerged from an M--theory compactification of the form
$\text{G\"{o}del}\times S^{2}\times CY_{3}$, which was interpreted as
coming from the backreaction of M2--branes wrapping the $S^{2}$.

In this note, our aim is to improve our understanding of these vacua
by their explicit realisation in string theory and to study their
supersymmetry properties (at the level of supergravity). In order to do so, we will make use of the
standard brane intersection rules for building supergravity solutions,
and the application of a form of T--duality at the level of the 
ten--dimensional theories. Our starting point will be the IIB setup
obtained by T--dualizing the $D=4$ extremal dyonic black string, which
has been widely studied in the past \cite{Duff:1995yh, Horowitz:1996ay, Boonstra:1997dy, Boonstra:1998yu}. We will then show that by adding
a plane wave or a monopole and T--dualizing along a fiber coordinate,
it is possible to obtain backgrounds in which a maximally symmetric
subspace becomes warped or squashed.

The plan of this paper is as follows. In
section~\ref{sec:from-d1d5m-wave} we state our main results; we
explicitly build the squashed/warped solutions in IIA supergravity and provide an
interpretation in terms of D--branes with monopoles and/or waves. We
then discuss the supersymmetry properties of the solutions in section~\ref{sec:susy-properties}
by direct computation of the associated Killing spinors and comment on specific limits of the
deformation parameter. Finally we present our conclusions and provide
some possible directions into future research.
\section{T--duality for D1/D5/monopole/plane wave backgrounds}
\label{sec:from-d1d5m-wave}
\subsubsection*{Main result}
In this section we derive our main result. By T--dualizing the $D=4$
extremal dyonic black string solution we can construct a type II string
setup with metric
\begin{equation}
  \di s^2 = \mathcal{M}_3 + \mathcal{E}^3 + T^4 \, ,  
\end{equation}
where $\mathcal{M}_3$ is either $\AdS_3$ or a \emph{warped
  anti--de Sitter space} with radius $R$ and warping parameter $\sinh
\Theta_w$, and $\mathcal{E}_3$ is either a three--sphere $S^3(R)$ or a
\emph{squashed three--sphere }with radius $R$ and squashing parameter
$\sinh \Theta_m$.
\subsubsection*{The extremal dyonic black string}
Consider the type IIB setup obtained as the superposition of a D1/D5
system with a magnetic monopole and a plane wave. This was already
described in~\cite{Tseytlin:1996bh} as the T--dual to the $D=4$
extremal dyonic black string\footnote{The intersection of an \textsc{ns--ns} 1
  -brane and a \textsc{ns--ns} 5-brane generates a dyonic black string in
  ten-dimensions that when reduced to six yields the dyonic black
  string geometry.}. The metric, dilaton and RR three--form field are 
\begin{multline}
  \di s^2 = H_1^{1/2} H_5^{1/2} \left( H_1^{-1} H_5^{-1} \left( \di u
      \di v + K \di u^2 \right) + H_5^{-1} \left( \di y_1^2 + \dots \di
      y_4^2 \right) + \right. \\ + \left. V^{-1} \left( \di \psi + A_i
     \di x^i \right)^2 + V \left( \di x_1^2
      + \dots \di x_3^2 \right)\right) \, ,
\end{multline}
\begin{align}
  e^{2\phi} = H_1^{-1} H_5 \, ,&& F_{[3]} = H_1^{-1} \di t \wedge \di u \wedge \di v - B_i \di x^i \wedge \di \psi \, ,
\end{align}
where $H_1(\mathbf{x}), H_5 (\mathbf{x}), K(\mathbf{x}),
V(\mathbf{x}), A_i(\mathbf{x}), B_i(\mathbf{x})$ are harmonic
functions of the transverse coordinates $x_i, i=1,2,3$ and
\begin{align}
  \di B = - * \di H_5 \, , && \di A = - * \di V \, .
\end{align}
Passing to spherical coordinates $(r, \theta, \phi)$ for $x^i$ and
taking the $r \to 0$ limit, the harmonic functions take the form
\begin{align}
  H_1 = \frac{Q_1}{r} \, , &&  H_5 = \frac{Q_5}{r} \, , && K = \frac{Q_w}{r} \, , && V = \frac{Q_m}{r} \, , 
\end{align}
and in a suitable coordinate system the field configuration
becomes 
\begin{multline}
  \di s^2 = Q_m Q_1^{1/2} Q_5^{1/2} \left( - \di \tau^2 + \di \omega^2
    + Q_w \di \sigma^2 + 2{Q_w}^{1/2} \sinh \omega \di \sigma \di \tau \right) + \\
  + Q_m Q_1^{1/2} Q_5^{1/2} \left( \di \theta^2 + \di \phi^2 + \di
    \psi^2 + 2 \cos \theta \di \psi \di \phi \right) + Q_1^{1/2}
  Q_5^{-1/2} \left( \di y_1^2 + \dots + \di y_4^2 \right) \, ,
\end{multline}
\begin{align}
  e^{2\phi} = Q_1^{-1} Q_5 \, ,&& F_{[3]} = Q_m Q_1^{1/2} Q_5^{1/2}
  \left( \cosh \omega \di \tau \wedge \di \omega \wedge \di \sigma +
    \sin \theta \di \phi \wedge \di \psi \wedge \di \theta \right) \,
  .
\end{align}
Note that the metric is still $\AdS_3 \times S^3 \times T^4$,
just like in the near--horizon limit of the more standard $D1/D5$
system\footnote{Here the limit $Q_m \rightarrow 0$ is singular, since in this configuration 
we only have 3 transverse coordinates}. Some of the variables are periodic by construction. Moreover,
one can impose a discrete identification in the anti--de Sitter part,
leading to a BTZ black hole~\cite{Banados:1992wn}. In detail we have
the following periodicities: 
\begin{equation}
  \begin{cases}
    \psi = \psi + 4 \pi \, ,\\
    \sigma = \sigma + 4 \pi \, ,\\
    y_i = y_i + 2 \pi \, .
  \end{cases}
\end{equation}
This is not unlike the BTZ identification, which strictly speaking differs from the black string geometry, but which is nevertheless well understood. Keeping this into account we can introduce a new pair of $4\pi$--periodic variables $\alpha $ and $\beta$: 
\begin{align}
  \psi = \alpha + 2 y_1 && \sigma = \beta + 2 y_2 \, ,  
\end{align}
Notice that the $y_{i}$ coordinates stop describing an external (to $\AdS_{3}\times S^{3}$) torus when we introduce the coordinates $\alpha$ and $\beta$, which are linear combinations of the $y_i$ and angular coordinates in $\AdS$ and the sphere respectively. We now rewrite the metric as: 
\begin{multline}
  \label{eq:IIB-metric}
  \di s^2 = R^2 \left[ \di \omega^2 - \cosh^2 \omega \di \tau^2 +
    \frac{1}{\cosh^2 \Theta_w} \left( \di \beta + \sinh \omega \di
      \tau \right)^2 \right]  \\
  + R^2 \left[ \di \theta^2 + \sin^2 \theta \di \phi^2 +
    \frac{1}{\cosh^2 \Theta_m} \left(\di \alpha + \cos \theta \di \phi
    \right)^2 \right] +   \frac{R^2}{\sinh^2 \Theta_m} \left( \di y_3^2 + \di y_4^2 \right)
 \\
  +\left( \di z_w + R \tanh \Theta_w \left( \di \beta + \sinh \omega
      \di \tau \right) \right)^2 + \left( \di z_m + R \tanh \Theta_m
    \left( \di \alpha + \cos \theta \di \phi \right) \right)^2   \, ,
\end{multline}
where the parameters $R$, $\Theta_m$ and $\Theta_w$ are related to the charges by
\begin{align}
  R^2 = Q_m \sqrt{Q_1 Q_5} \, , && \sinh^2 \Theta_m = 4 Q_m Q_5 \, , && \sinh^2 \Theta_w = 4 Q_w Q_m Q_5 \, ,
\end{align}
and
\begin{align}
  z_m = \frac{2R}{\tanh \Theta_m}  y_1 \, , &&   z_w = \frac{2R}{\tanh \Theta_w} \tilde{y}_2 \, .
\end{align}
where $\tilde{y}_{2}=\sqrt{Q_{w}}y_{2}$.

\subsubsection*{Type IIA backgrounds}
Up to this point we have only obtained a rewriting of the background
fields. Now comes the main point in our construction. Both the
$\AdS_3$ and $S^3 $ geometries can be understood as Hopf fibrations
(respectively of $\AdS_2$ and $S^2$), and performing a T-duality in
the direction of the fiber can undo the structure. The only technical
problem that arises in this situation is related to the presence of
the Ramond--Ramond fields that are not considered in the usual Buscher
transformations~\cite{Buscher:1987sk}. This situation has already been
studied in the literature (see
\emph{e.g.}~\cite{Bergshoeff:1995as, Bergshoeff:1996ui, Duff:1998cr, Hassan:1999bv, Cvetic:1999zs, Benichou:2008it}). Nevertheless,
it is instructive to consider a possible approach in detail. Starting
with the type IIB background in Eq.~(\ref{eq:IIB-metric}) we obtain a
type IIA T--dual background by the following set of transformations
(described in more detail in Appendix~\ref{sec:t-duality-with-RR}):
\begin{enumerate}
\item Reduce the type IIB background to nine dimensions in the
  direction $z_i$
\item Rewrite the nine--dimensional IIB fields in terms of
  nine--dimensional IIA fields (in nine dimensions type IIA and IIB
  are the same)
\item Oxidate to ten dimensional type IIA introducing the variables
  $\zeta_i$, T--dual of $z_i$.
\end{enumerate}

\paragraph{The Squashed Sphere. }

As an example we will now apply this construction by performing a
T--duality in the $z_m$ direction. If we single out the
coordinates in the sphere component, the fields read
\begin{subequations}
  \begin{align}
    \begin{split}
      \di s_{10}^2 &= \AdS_3 [R] + T^3 + R^2 \left[ \di \theta^2
        + \sin^2 \theta \di \phi^2 + \frac{1}{\cosh^2 \Theta_m}
        \left(
          \di \alpha + \cos \theta  \di \phi \right)^2 \right]  \\
      & \quad +\left[ \di z_m + R \tanh \Theta_m \left( \di \alpha + \cos
          \theta \di \phi \right) \right]^2
    \end{split} \\
    F_3 &= \vol_{\AdS} + R^2 \sin \theta \di \theta \land \di \phi
    \land \di \alpha + R \tanh \Theta_m \sin \theta \di \theta \land
    \di \phi \land \di z ,
  \end{align}
\end{subequations}
and we can perform a Kaluza-Klein reduction on $z_m$ and go to nine
dimensions. The metric reads:
\begin{equation}
  \di s_{9}^2 = \AdS_3 [R] + T^3 + R^2 \left[ \di \theta^2
    + \sin^2 \theta \di \phi^2 + \frac{1}{\cosh^2  \Theta_m} \left( \di \alpha + \cos \theta
      \di \phi \right)^2 \right] ,
\end{equation}
and the gauge fields are obtained from:
\begin{equation}
  F_3 = F^{(3)}_3 + F_2^{(3)} \land \left( \di z_m + A \right) ,   
\end{equation}
where $F_m^{(n)}$ is the $m$-form obtained from the reduction of a
$n$-form and $A$ is the one-form
\begin{equation}
  A = R \tanh \Theta_m \left( \di  \alpha + \cos \theta \di \phi \right) .
\end{equation}
Explicitly, adding the extra Kaluza-Klein two-form:
\begin{align}
  F^{(3)}_3 &=  \vol_{\AdS} + \frac{R^2}{\cosh^2 \Theta_m} \sin \theta \di \theta \land \di \phi \land \di \alpha \\
  F_2^{(3)} &= R \tanh \Theta_m \sin \theta \di \theta \land \di \phi \\
  F_2^{(g)} &= \di A = R \tanh \Theta_m \sin \theta \di \theta \land \di \phi  .
\end{align}
Now, let us perform a T-duality to go to type IIA. Given that there is
only one supergravity theory in nine dimensions, the fields keep their
expressions but the interpretation changes according to
Tab.~\ref{tab:T-duality}: $F_3^{(3)} $ now comes from the reduction of
a four-form in ten dimensions, $F_2^{(2)} $ from a two-form and
$F_2^{(g)}$ is now obtained as the result of the reduction of the
Kalb-Ramond field:
\begin{align}
  F_3^{(4)} = F_3^{(3)} && F_2^{(2)} = F_2^{(3)} && F_2^{(B)} = F_3^{(g)} .
\end{align}
We can oxidise back to ten dimensions and get a IIA background:
\begin{subequations}
  \begin{align}
    \di s_{10}^2 &= \AdS_3 [R] + T^3 + R^2 \left[ \di \theta^2 +
      \sin^2 \theta \di \phi^2 + \frac{1}{\cosh^2 \Theta_m} \left(
        \di \alpha + \cos \theta
        \di \phi \right)^2 \right] + \di \zeta^2_m \\
    F_4 &= F_3^{(4)} \land \di \zeta_m = \left[ \vol_{\AdS} +
      \frac{R^2}{\cosh^2 \Theta_m} \sin \theta \di \theta \land \di
      \phi \land \di \alpha \right] \land \di \zeta_m \\
    F_2 &= F_2^{(2)} = R \tanh \Theta_m \sin \theta \di \theta \land \di \phi \\
    H_3 &= F_2^{B} \land \di \zeta_m = R \tanh \Theta_m \sin \theta \di
    \theta \land \di \phi \land \di \zeta_m \, .
  \end{align}
\end{subequations}

Even though the squashed sphere is not a group manifold, we can still
use techniques borrowed from group theory. In particular, we can derive
that the isometry group for this part of the metric is $SU(2) \times
U(1)$ and that the spectrum of the scalar Laplacian is
\begin{equation}
  \triangle_{\Theta_m} Y_{lj} = \frac{1}{R^2} \left\{ l \left( l + 1 \right) + \sinh^2 \Theta_m \, j^2 \right\} Y_{lj} \hspace{2em} l = 0, 1/2, 1, \dots ; \ j = -l, \dots, l \, ,
\end{equation}
where $Y_{lj}$ are the usual three--dimensional spherical
harmonics. More details can be found in
Appendix~\ref{sec:lapl-squash-sphere}.

 
It is worthwhile to emphasize that by construction $\alpha $ is $4
\pi$-periodic and  the geometry is the one of a respectable
squashed three-sphere. 

A similar construction was considered in~\cite{Duff:1998cr}. In that
case, though, the authors started with an $\AdS_3 \times S^3$ geometry
supported by \emph{both} \textsc{rr} and \textsc{ns--ns} fields that
was then reduced on one of the sphere isometries, yielding a Lens
space $S^3/\setZ_p$ or a squashed version, where $p$ and the squashing
depend on the values of the charges. This is clearly an orbifold of
the solutions above. Our construction starts with a different
background (which also includes a monopole and a plane wave) and RR
fluxes and we consider T-duality along an extra-dimension which is a
linear combination of the Hopf fiber coordinate on the $S^3$ (or
$\AdS_3$) and a torus direction. To be more specific, consider the $S^3 \times S^1
$ part. The geometry can be understood as the fibration
  \begin{equation}
    \begin{CD}
      S^1 \times S^1 @>>> S^3 \times S^1 \\
      @.      @VVV\\
      {} @. S^2
    \end{CD}
  \end{equation}
where one of the directions in the torus fibration is the Hopf fiber in $S^3$. A $S_1$ sub-bundle $A$ of the torus, obtained as a rational
linear combination of the two directions, describes a fibration
  \begin{equation}
    \begin{CD}
      A @>>> S^3 \times S^1 \\
      @.      @VVV\\
      {} @. \SqS^3
    \end{CD}
  \end{equation}
where the squashing parameter depends on the coefficients in the linear combination.  Note that just like in the pure $S^3 $ case described in ~\cite{Duff:1998cr}, since we only have RR fields, performing T--duality in the $A$ direction will "unwind" the fiber and lead to a geometry which is the direct product $\SqS^3 \times S^1$. The same can be applied to the $\AdS_3 \times S^1 $ part.

Finally, given that our starting geometry is supported by RR fluxes only, our construction is not suited to describe the exact
CFT backgrounds of~\cite{Israel:2004vv, Detournay:2005fz, Orlando:2006cc}. Since these
latter are based on Wess--Zumino--Witten models, they have \textsc{ns--ns}
fields instead of \textsc{r--r} fields a T--duality in the $z_m$ direction
would not trivialize the Hopf fibration.


\paragraph{Warped AdS.}
In principle these constructions can be extended to other group
manifold geometries (\emph{e.g.} the obvious choice leading to a
squashed $\AdS_3$) but in any case one should start from a
configuration with \textsc{rr} fields, since the absence of
\textsc{ns--ns} antisymmetric fields is the key ingredient for the
trivialization of the fiber bundle. More general geometries can be
obtained by starting with a mixed \textsc{rr}-\textsc{ns--ns}
configuration.  Starting with the background in
Eq.~\eqref{eq:IIB-metric}, the construction can be performed in two
more ways:
\begin{enumerate}
\item Performing the T--duality on the coordinate $z_w$. This leads to
  a type IIA background whose metric is $\WAdS_3 \times S^3 \times
  T^4$, sustained by a four--form flux, a two--form flux and a
  Kalb--Ramond field. Explicitly:
  \begin{subequations}
    \label{eq:WAdS3-S3-T4}
    \begin{align}
      \di s^2 &= R^2 \left[ \di \omega^2 - \cosh^2 \omega \di \tau^2 +
        \frac{1}{\cosh^2 \Theta_w} \left( \di \beta + \sinh \omega \di
          \tau \right)^2 \right] + \di \zeta_w^2 + S^3 [R] + T^3 \, , \\
      F_4 &= \left[ R^2 \ \vol_{\mathrm{S}} + \frac{R^2}{\cosh^2
          \Theta_w} \cosh \omega \di \omega \land \di \tau \land \di
        \beta
      \right] \land \di \zeta_w \, , \\
      F_2 &= R \tanh \Theta_w \cosh \omega \di \omega \land \di \tau \, ,\\
      H_3 &= R \tanh \Theta_w \cosh \omega \di \omega \land \di \tau
      \land \di \zeta_w \, .
    \end{align}
  \end{subequations}
\item Performing two T--dualities in both $z_m$ and $z_w$, which leads
  to a type IIB background with metric $\WAdS_3 \times \SqS^3 \times
  T^4$. In this case, following Appendix~\ref{sec:t-duality-with-RR},
  we find that the metric is sustained by a five--form flux, a
  three--form flux and a Kalb--Ramond field. Explicitly:
  \begin{subequations}
    \begin{align}
      \begin{split}
        ds^{2} &= R^2 \left[\di \theta^2 + \sin{\theta}^2 \di \phi^2 + \frac{1}{\cosh^2 \Theta_m} \left(\di \alpha + \cos{\theta} \di\phi \right)^2 \right] \\
        & \quad + R^2 \left[ -\cosh^2{\omega} \di \tau^2 + \di \omega^2 + \frac{1}{\cosh^2 \Theta_w} \left(\di \beta + \sinh{\omega} \di \tau \right)^2 \right] + d\zeta_w^2 + d\zeta_m^2 + T^2 \, ,
      \end{split} \\
      F_{5} &= R^2 \left[ \frac{1}{\cosh^2 \Theta_m} \sin{\theta} \di\theta\wedge \di\phi \wedge \di \alpha  +\frac{1}{\cosh^2 \Theta_w} \cosh{\omega} \di \tau \wedge \di \omega \wedge \di \beta\right] \wedge \di \zeta_w \wedge \di \zeta_m \, , \\
      F_{3} &= - R \tanh \Theta_w \cosh{\omega} \di \tau \wedge \di \omega \wedge \di \zeta_m + R \tanh \Theta_m \sin{\theta} \di \theta \wedge \di \phi \wedge \di \zeta_w \, , \\
      H_{3} &= - R \tanh \Theta_w \cosh{\omega} \di \tau \wedge \di \omega \wedge \di \zeta_w + R \tanh \Theta_m \sin{\theta} \di \theta \wedge \di \phi \wedge \di \zeta_m \, .
    \end{align}
    \label{twotdualities}
  \end{subequations}
\end{enumerate}

\section{Supersymmetry Properties}
\label{sec:susy-properties}

T--duality transformations can break supersymmetries preserved by
D--brane solutions. This has already been observed in the context of
Hopf T-dualities on six-dimensional supergravity backgrounds of the
form $\AdS_{3} \times S^{3}$~\cite{Duff:1998cr}. The phenomenon is
akin to the breaking of supersymmetry by compactification on a circle,
so the resulting supersymmetries will be those that survive the circle
compactification.

In the following we compute the explicit expressions of the Killing
spinors for the geometries obtained above, and verify that under T--duality, the IIA backgrounds
with squashed/warped spacetimes \emph{preserve $1/4$ of the
  supersymmetries} of the original D1/D5 background. We also show that
for specific values of the deformation parameter some supersymmetries
are restored and that generically, IIB backgrounds containing both
warped AdS and squashed spheres preserve no supersymmetry.

\subsection{$\AdS_{3}\times S^{3}\times T^4$}

We start by calculating the Killing spinors in the $\AdS_3\times S^3
\times T^4$ background\footnote{Killing spinors for backgrounds of the
  form $\AdS_{p}\times S^{q} $ have been computed in horospherical
  coordinates in~\cite{Lu:1998nu}.  }. The details can be found in
appendix~\ref{a}, here we will just quote the main results. The
Killing spinors read:
\begin{equation}
  \epsilon=e^{-\frac{\omega}{2}\gamma^{02}\otimes\mathcal{P_{+}}}e^{\frac{t}{2}\gamma^{12}\otimes\mathcal{P_{+}}}e^{\frac{\sigma}{2}\gamma^{01}\otimes\mathcal{P_{-}}}
  e^{\frac{\theta}{2}\gamma^{45}\otimes\mathcal{P_{+}}}e^{\frac{\psi}{2}\gamma^{34}\otimes\mathcal{P_{+}}}e^{\frac{-\phi}{2}\gamma^{34}\otimes\mathcal{P_{-}}}\epsilon_0
\end{equation}
where $\mathcal{P}_{\pm}=\frac{\mathbb{1}\pm\sigma^{1}}{2}$ are
projection operators acting on the doublet $\epsilon =
\left( \begin{smallmatrix}\epsilon_1\\
    \epsilon_2 \end{smallmatrix}\right) $ and the constant spinors
$\epsilon_0$ are also complex Weyl spinors arising from the
integration constants. These spinors are contrained as to satisfy the
standard $T^4$ projection
\begin{align}
  \left(\gamma^{012345}+1\right)\epsilon_0=0.
\end{align}
which implies the preservation of $1/2$ of the supersymmetries
(\emph{i.e.} $16$ supersymmetries remain unbroken). 

Further insights can be obtained by looking at the explicit
expressions of the Killing spinors. Half of them depend on the fiber
coordinates $\sigma, \phi$: we conclude \emph{a priori} that under T--duality, the spinors associated to those supersymmetries are
broken and the IIA background will preserve 1/4 of the original
supersymmetry.
\subsection{Warped and Squashed Backgrounds}

We now compute the Killing spinors for the T-dual backgrounds
explicitly. For the technical details we refer the reader to
appendix~\ref{iia}. We take $\epsilon$ to be Majorana but not
Weyl.

If the T-duality is performed along the sphere fiber $\alpha$, one
obtains
$\epsilon=\epsilon_1(\omega,\tau,\sigma)\epsilon_2(\theta,\phi,\alpha)\epsilon_0$, where $\epsilon_0$ is a constant Majorana spinor and
\begin{align}
  \epsilon_{1}(\omega,\tau,\sigma)&=\exp\left\{-\frac{\omega}{2}\gamma^{02}\mathcal{Q}_{+}\right\}\exp\left\{\frac{\tau}{2}\gamma^{12}\mathcal{Q}_{+}\right\}\exp\left\{\frac{\sigma}{2}\gamma^{01}\mathcal{Q}_{-}\right\} \\
  \begin{split}
    \epsilon_{2}(\theta,\phi,\alpha) &= \exp\left\{\frac{\theta}{2}\gamma^{0123}\mathcal{Q}_{+}\right\}\left[\exp\left\{ \tanh^2{\Theta_m}\cos{\theta}\left(\frac{\phi}{2}\right)\gamma^{34} \right\}\mathcal{Q}_{-} \right.   \\
    & \quad +\left.\exp\left\{\frac{\phi}{2}\gamma^{34}
      \right\}\mathcal{Q}_{+}\right]\exp\left\{-\sech^2\Theta_{m}\frac{\alpha}{2}\gamma^{34}\mathcal{Q}_{-}\right\}
  \end{split}
\end{align}
and $\mathcal{Q}_{\pm}=\frac{\mathbb{1}\pm\gamma^{9}}{2}$. The Killing
spinor is also subject to the constraint:
\begin{equation}
  \gamma^{34}\mathcal{Q}_{+}\Gamma^{11}\epsilon=0 
\end{equation}
which turns into a constraint on $\epsilon_0$. Writing $\epsilon=\epsilon_{+}+\epsilon_{-}$ with
$\mathcal{Q}_{\pm}\epsilon_{\pm}=\pm\epsilon_{\pm}$ then one can
verify that only the supersymmetries associated to the $\epsilon_{-}$
spinors are preserved after T--duality. \emph{The solution is
  $1/4$-BPS}.

When the deformation parameter is such that
$\sech{\Theta_m}=0$, there are no spinors depending on the T--dual
coordinate and supersymmetry is restored to $1/2$-BPS. This
corresponds to the case in which the $S^{3}$ becomes $S^2 \times S^1$ \cite{Orlando:2006cc, Orlando:2005vt}.

\bigskip

In the other case, when the T--duality is done along the $\AdS$ fiber
$\beta$, one expresses the Killing spinor as
$\epsilon=\epsilon_1(\theta, \psi, \phi)
\epsilon_2(\omega,\tau,\beta)\epsilon_0$, with
\begin{align}
  \epsilon_{1}(\theta,\psi,\phi)&=\exp\left\{\frac{\theta}{2}\gamma^{45}\mathcal{Q}_{+}\right\}\exp\left\{\frac{\psi}{2}\gamma^{34}\mathcal{Q}_{+}\right\}\exp\left\{-\frac{\phi}{2}\gamma^{34}\mathcal{Q}_{-}\right\} \\
  \begin{split}
    \epsilon_{2}(\omega,\tau,\beta) &= \exp\left\{-\frac{\omega}{2}\gamma^{1345}\mathcal{Q}_{-}\right\}\left[\exp\left\{ -\tanh{\Theta_w}^2\sinh{\omega}\left(\frac{\tau}{2}\right)\gamma^{01} \right\}\mathcal{Q}_{+} \right. \\
    & \quad +\left.\exp\left\{\frac{\tau}{2}\gamma^{0345}
      \right\}\mathcal{Q}_{-}\right]\exp\left\{-\sech^2\Theta_w\frac{\beta}{2}\gamma^{01}\mathcal{Q}_{+}\right\}
  \end{split}
\end{align}
with the spinor satisfying the constraint
\begin{equation}
\gamma^{01}\mathcal{Q}_{+}\Gamma^{11}\epsilon=0 
\end{equation}
which once more, turns into a constraint on $\epsilon_0$. For $\epsilon=\epsilon_{+}+\epsilon_{-}$, only the supersymmetries
associated to the $\epsilon_{+}$ spinors will be preserved after
T--duality \emph{and the solution is again $1/4$-BPS}.

Just as before, in the special case when the deformation parameter is
such that $\sech{\Theta_w}=0$, $\AdS_{3}$ becomes $\AdS_{2}\times
S^1$, there are no spinors depending on the T-dual coordinate, and
supersymmetry is restored to $1/2$-BPS.
\subsection{$\WAdS_{3}\times \SqS^{3}\times T^4$}
Instead of determining the expressions for the Killing spinors of the
IIB background obtained after two T--dualities in which both the
sphere and the anti-de Sitter space are squashed/warped, we can
directly argue that all supersymmetries must be in general broken.

Let $\epsilon$ be a Weyl complex spinor in 10 dimensions and let
$\mathcal{T}_\pm = \frac{\mathbb{1}\pm\gamma^{67}}{2}$ be a projection
operators. Decomposing $\epsilon=\epsilon_{+}+\epsilon_{-}$ with
$\mathcal{T}_{\pm}\epsilon_{\pm}=\pm\epsilon_{\pm}$, one can check
that for this background the following conditions need to be held.
\begin{itemize}
\item From the vanishing of the dilatino variation
  \begin{equation}
    \tanh{\Theta_{w}}\gamma^{01}\epsilon_{-}=0 \qquad \qquad \tanh{\Theta_{m}}\gamma^{34}\epsilon_{+}=0 \, .
  \end{equation}
 \item The Killing spinor satisfies $\Gamma^{11} \epsilon=\epsilon$.
\end{itemize}
Imposing these projections necessarily yields the condition
\begin{equation}
  \tanh{\Theta_m} = - i \tanh{\Theta_w}
\end{equation}
for supersymmetry to be preserved. Hence, for generic values of the
deformations \emph{all the associated supersymmetries must be broken}.
However, it is possible to see that there is restauration of $1/2$-BPS
in the special case of $\AdS_2 \times S^2 \times T^6 $ background.

\section{Conclusions}
\label{sec:conclusions}

In this note we have explicitly constructed string theory backgrounds
which include three--dimensional squashed/warped spheres/anti-de
Sitter spaces. They can be seen as T--dual brane
intersections of D1 and D5 systems with monopole and/or plane
waves. The deformation parameters of the squashed/warped spaces are
 interpreted in terms of the charges of the original background.

We also studied the supersymmetry properties of these backgrounds by
explicitly calculating the Killing spinors. Under T--duality, IIA
backgrounds containing $\SqS^3$ or $\WAdS_3$ preserve eight
supersymmetries for generic values of the deformation parameter. In
the special case of infinite deformation parameter
$\cosh{\Theta_{m,w}} \to \infty$, some supersymmetries are restored
and the backgrounds preserve sixteen supersymmetries. The same
arguments can be applied to the IIB background in which both the
sphere and the $\AdS$ space are squashed/warped, but for generic
values of the deformation parameter, no supersymmetry is preserved. 

It should be remarked that the previous results were obtained in the context of
supergravity. It is well-known that spacetime supersymmetries that are
manifest in some string background might very well be hidden in their
T--dual~\cite{Bakas:1995hc,Sfetsos:1995ac}. There are examples in the literature in which
supersymmetry that seemed to be destroyed by duality could actually be
restored by a non--local
realization~\cite{Hassan:1995je,Ricci:2007eq}. To clarify these
issues, it would be necessary to study the precise T--duality here addressed at the level of
the GS action for a IIB $\AdS_3\times S^3 \times T^4$
background~\cite{Babichenko:2009dk}. An analysis from the perspective
of the dual CFT would also be enlightening as all worldsheet and
spacetime supersymmetries should remain symmetries of the underlying
CFT. We leave these and related questions as future work.

The same construction works for any metric that posses an $S^1$
fibration and formally also for $S^3$ fibers, even though in this
case, some modifications would need to be included (in particular in
order to generalize the T--duality to non--Abelian fields). Another
possibility would be to follow the procedure along the time-like $S^1$
fiber in $\AdS_3$, which would lead to a hyperbolic plane ($H^2$)
geometry in type $II^{*}$ backgrounds with negative kinetic terms.

It would be interesting to use these solutions to compute holographic
data, thus formulating in a more precise way the duality chain
in~(\ref{dualitychainhopf}). For the case in which the $\AdS$ space
becomes warped after T--dualizing, an appropriate embedding in an
asymptotically AdS background would be required, following the
procedure in~\cite{Levi:2009az}). We believe the geometries here presented constitute a first step 
towards attempting to uncover the microscopic duals of these backgrounds. This issue is currently under investigation. 

Another possibility that we have not fully explored in the present
note consists in T--dualizing the dyonic black string background along
a direction that mixes the two initial coordinates $\psi $ and
$\sigma$. The resulting geometry (in type IIA) is described by the
metric
\begin{multline}
  \di s^2 = R^2 \left[ \di \omega^2 - \cosh^2 \omega \di \tau^2 +
    \frac{1}{\cosh^2 \Theta_w} \left( \di \beta + \sinh \omega \di
      \tau \right)^2 \right] + \\
  + R^2 \left[ \di \theta^2 + \sin^2 \theta \di \phi^2 +
    \frac{1}{\cosh^2 \Theta_m} \left(\di \alpha + \cos \theta \di \phi
    \right)^2 \right] + \\ + 2 R^2 \tanh \Theta_w \tanh \Theta_m \left( \di \beta + \sinh \omega
    \di \tau \right) \left( \di
    \alpha + \cos \theta \di \phi \right) + T^4 \, , 
\end{multline}
which can be reduced to any of the previously discussed configurations, $\WAdS_3 \times S^3 $ and $\AdS_3
\times \SqS^3$, by setting the relevant charges to zero.

\subsection*{Acknowledgements}

We are indebted to Jos\'{e} Figueroa-O'Farrill, Shigeki Sugimoto,
Kostas Skenderis for enlightening discussions and especially to Arkady
Tseytlin for comments and his careful revision of the manuscript. We
would furthermore like to thank the participants of the IPMU string
theory group meetings for directing our attention to the question
investigated in this note. L.I.U. would like to thank IPMU for
hospitality during the initial stages of this project.\\
The research of D.O. was supported by the World Premier International
Research Center Initiative (WPI Initiative), MEXT, Japan.  L.I.U. is
supported by a STFC Postdoctoral Research Fellowship.

\bibliography{Biblia}

\appendix

\section{Group theoretical description of the geometry}
\label{sec:lapl-squash-sphere}

\paragraph{Isometries.}
In order to understand the isometries of the squashed and warped
spaces it is convenient to describe their geometry in algebraic
terms. This is possible because both the three--sphere and $\AdS_3$
are group manifolds, respectively for $SU(2)$ and $SL(2, \setR)$. It
follows that their line elements can be written as
\begin{equation}
  \di s^2 = R^2 \Tr[ g^{-1} \di g g^{-1} \di g] = R^2 \sum_{a} J^a \otimes J^a \, ,  
\end{equation}
where $g \in G$ is a general element of the group, $J^a = \Tr [ t^a g^{-1}
\di g ]$ and $t^a$ are the generators of the Lie algebra ($su(2) $ or
$sl_2$). The effect of the T--duality amounts to adding an extra term
to the metric, proportional to $\left( J^3 \right)^2$ (in the $sl_2$
case, $J^3 $ is the hyperbolic generator). The squashed/warped metric
is  written as
\begin{equation}
  \di s^2[\Theta] = R^2 \left( J^1 \otimes J^1 + J^2 \otimes J^2 + \frac{1}{\cosh^2 \Theta} J^3 \otimes J^3 \right) \, .  
\end{equation}
The initial group manifold has $G \times G$ isometry, generated by
$J^a = \Tr [ t^a g^{-1} \di g]$ and $\bar J^a = \Tr [ t^a \di g
g^{-1}]$, but only part of this symmetry remains after the
T--duality. To be precise, while the $\bar J^a $ generators are
preserved, since they commute with the current $J^3$, both $J^1$ and
$J^2 $ are not Killing vectors anymore, as one can verify with a
direct calculation of the Lie derivative of the metric:
\begin{gather}
  \mathcal{L}_{J^1} [ \di s^2[\Theta] ] = 2 R^2 \tanh \Theta \, J^2 \otimes J^3 \, , \\
  \mathcal{L}_{J^2} [ \di s^2[\Theta] ] = 2 R^2 \tanh \Theta \, J^3 \otimes J^1 \, .
\end{gather}
The resulting isometry group is $G \times U(1)$.

\paragraph{Scalar Laplacian.}
As an application of this description, let us derive the spectrum of
the scalar Laplacian on the squashed/warped geometries. Consider in
particular the squashed three--sphere $\SqS^3(R,\sinh \Theta_m)$.
Let $E^a$ be the basis vectors in $S^3$, such that $\braket{J^a, E_b}
= \delta^a_{\phantom{a}b}$. The drei--bein $\set{\theta^a}$ and the
dual basis vectors $\set{e_a}$ on $\SqS^3(R, \sinh \Theta_m)$, defined
by
\begin{align}
  \di s^2[\SqS^3(R,\sinh \Theta_m)] = \sum_{a=1}^3 \theta^a \otimes \theta^a \, , &&      \braket{\theta^a, e_b} = \delta^a_{\phantom{a}b} \, ,
\end{align}
can be expressed as
\begin{align}
  \theta^1 = R J^1 \, , && \theta^2 = R J^2 \, , && \theta^3 = \frac{R}{\cosh \Theta_m} J^3 \, , \\
  e_1 = \frac{1}{R} E_1 \, ,&& e_2 = \frac{1}{R} E_2 \, ,&& e_3 = \frac{\cosh \Theta_m}{R} E_3 \, .
\end{align}
By definition, the connection one--form $\omega$ is given by
\begin{equation}
  \di \theta^a = - \omega^a_{\phantom{a}b} \wedge \theta^b \, ,  
\end{equation}
and since
\begin{equation}
  \di J^a = \frac{1}{2} \epsilon_{abc} J^b \wedge J^c \, ,  
\end{equation}
it is clear that $\omega^a_{\phantom{a}b} $ is proportional to
$\epsilon_{abc}\theta^c$ and hence $\braket{\omega^a_{\phantom{a}b},
  e_a}=0$. We have now all the ingredients to express the scalar
Laplacian operator on $\SqS$ in terms of operators on the
three--sphere:
\begin{multline}
  \triangle_{\Theta_m} f = - e_a e_a f + \braket{\omega^a_{\phantom{a}b}, e_a} e_b f = \frac{1}{R^2} \left\{  \left( E_1^2 + E_2^2 + E_3^2 \right) + \sinh^2 \Theta_m E_3^2 \right\} =\\= \frac{1}{R^2} \left( \triangle_0 + \sinh^2 \Theta_m E_3^2 \right)  \, . 
\end{multline}
Now observe the following:
\begin{itemize}
\item $E_1^2+ E_2^2 + E_3^2 = \triangle_0$ is the Laplacian on the
  three--sphere. This has eigenvalues $l \left( l + 1 \right)$, $l \in
  0, 1/2, 1, \dots$
\item $E_3$ is the $z$ component of the angular momentum on $S^3$. It
  has eigenvalue $j, -l \le j \le l$.
\item Since the Laplacian on $S^3$ is also the Casimir of $SU(2)$,
  $\comm{\triangle_0,E_3}= 0$.
\end{itemize}
The operators $\triangle_{\Theta_m}$, $\triangle_0$ and $E_3$ are
commuting and admit the same eigenvectors. Moreover, the spectrum of
$\triangle_{\Theta_m}$ is given by the sum of the two
contributions. Explicitly, the eigenvalue equation reads
\begin{equation}
  \triangle_{\Theta_m} Y_{lj} = \frac{1}{R^2} \left\{ l \left( l + 1 \right) + \sinh^2 \Theta_m \, j^2 \right\} Y_{lj} \hspace{2em} l = 0, 1/2, 1, \dots ; \ j = -l, \dots, l \, ,
\end{equation}
where $Y_{lj}$ are the three--dimensional spherical harmonics.

In a similar way, one finds that the Laplacian spectrum on $\WAdS_3
(R, \sinh^2 \Theta_w)$ is given by the sum of the Laplacian of
$\AdS_3$ and an extra component $\sinh^2 \Theta_w / R^2 j^2$.

\section{T duality with \textsc{r--r} fields}
\label{sec:t-duality-with-RR}

In the II solutions we consider, the r\^ole of sustaining the geometry
is taken by \textsc{r--r} field strengths. In particular this means that
the usual Buscher rules~\cite{Buscher:1987sk} prove insufficient and
we are forced to follow a slightly more involved path to write
T-duals: derive two low-energy effective actions and explicitly write
down the transformations relating them (in this we will follow the
same procedure as in~\cite{Duff:1998us,Duff:1998cr}).

In ten dimensions, type IIA and IIB are related by a T-duality
transformation, stating that the former theory compactified on a
circle of radius $R$ is equivalent to the latter on a circle of radius
$1/R$. This means in particular that there is only one possible
nine-dimensional $N=2$ \textsc{sugra} action. The rules of T-duality
are easily obtained by explicitly writing the two low-energy
actions and identifying the corresponding terms.

For sake of clarity let us just consider the bosonic sector of both
theories. In~\cite{Lu:1995yn,Lu:1996ge} it was found that the IIA
action in nine dimensions is given by
\begin{multline}
  e^{-1} L_{IIA} = R - \frac{1}{2} (\partial \Phi)^2 - \frac{1}{2}(\partial
  \varPhi)^2 - \frac{1}{2} ({\mathcal{F}}\form{1}^{(12)})^2
  e^{\frac{3}{2}\Phi +\frac{\sqrt7}{2} \varPhi} + \\
  -\frac{1}{48} (F\form{4})^2 e^{\frac{1}{2}\Phi +\frac{3}{2\sqrt7} \varPhi}
  -\frac{1}{12} (F\form{3}^{(1)})^2 e^{-\Phi +\frac{1}{\sqrt7} \varPhi} -
  \frac{1}{12} (F\form{3}^{(2)})^2
  e^{\frac{1}{2}\Phi - \frac{5}{2\sqrt7} \varPhi} + \\
  -\frac{1}{4} (F\form{2}^{(12)})^2 e^{-\Phi - \frac{3}{\sqrt7}\varPhi}
  -\frac{1}{4} ({\mathcal{F}}\form{2}^{(1)})^2 e^{\frac{3}{2} \Phi
    +\frac{1}{2\sqrt7}\varPhi} - \frac{1}{4}
  ({\mathcal{F}}\form{2}^{(2)})^2
  e^{\frac{4}{\sqrt7} \varPhi} + \\
  -\frac{1}{2e} \tilde F\form{4} \land \tilde F\form{4} \land
  A\form{1}^{(12)} - \frac{1}{e} \tilde F\form{3}^{(1)} \land \tilde
  F\form{3}^{(2)} \land A\form{3}\ ,
\end{multline}
where $\Phi$ is the original dilaton, $\varPhi $ is a scalar measuring
the compact circle, defined by the reduction (in string frame)
\begin{equation}
  \di s^2 = e^{\Phi/2} \di s^2_{10} = e^{\Phi/2} \left( e^{-\varPhi/ \left(2 \sqrt{7} \right)} \di s^2_9 + e^{\sqrt{7} \varPhi /2} \left( \di z + \mathcal{A}\form{1} \right)^2 \right)    
\end{equation}
and $F\form{n}$ are $n$-form field strengths defined as
\begin{subequations}
  \begin{align}
    F\form{4} &=\tilde F\form{4} - \tilde F\form{3}^{(1)}\land
    \mathcal{A}\form{1}^{(1)} - \tilde F\form{3}^{(2)}\land
    \mathcal{A}\form{1}^{(2)} - \frac{1}{2} \tilde F\form{2}^{(12)}
    \land \mathcal{A}\form{1}^{(1)} \land  \mathcal{A}\form{1}^{(2)} \\
    F^{(1)}\form{3} &= \tilde F^{(1)}\form{3} - \tilde F\form{2}^{(12)} \land \mathcal{A}\form{1}^{(2)} \\
    F\form{3}^{(2)} &= \tilde F\form{3}^{(2)} + F\form{2}^{(12)} \land
    \mathcal{A}\form{1}^{(1)} - \mathcal{A}\form{0}^{(12)} \left( \tilde F^{(1)} \form{3} - F\form{2}^{(12)}\land \mathcal{A}\form{1}^{(2)} \right) \\
    F\form{2}^{(12)} &= \tilde F^{(12)} \form{2}\\
    \mathcal{F}\form{2}^{(1)} &= \mathcal{F}\form{2}^{(1)}
    + \mathcal{A}\form{0}^{(12)} \mathcal{F}\form{1}^{(2)} \\
    \mathcal{F}\form{2}^{(2)} &= \tilde{\mathcal{F}}\form{2}^{(2)} \\
    \mathcal{F}\form{1}^{(12)} &= \tilde{\mathcal{F}}\form{1}^{(12)}
    \ .
  \end{align}
\end{subequations}
In the same way, starting from the IIB action one obtains the
following nine-dimensional IIB Lagrangian:
\begin{multline}
  e^{-1} L_{IIB} = R - \frac{1}{2} (\partial \Phi)^2 -\frac{1}{2}
  (\partial
  \varPhi)^2 - \frac{1}{2} e^{2\Phi} (\partial  \chi)^2 + \\
  -\frac{1}{48} e^{-\frac{2}{\sqrt7} \varPhi} F\form{4}^2 -\frac{1}{12}
  e^{-\Phi+\frac{1}{\sqrt7}\varPhi} (F\form{3}^{({\textsc{ns}})})^2
  -\frac{1}{2} e^{\Phi
    +\frac{1}{\sqrt7} \varPhi} (F\form{3}^{({\textsc{r}})})^2 + \\
  -\frac{1}{4} e^{\frac{4}{\sqrt7} \varPhi} ({\mathcal{F}}\form{2})^2 -
  \frac{1}{4} e^{\Phi - \frac{3}{\sqrt7}\varPhi}
  (F\form{2}^{({\textsc{r}})})^2 -
  \frac{1}{4} e^{-\Phi - \frac{3}{\sqrt7} \varPhi} (F\form{2}^{({\textsc{ns}})})^2 + \\
  -\frac{1}{2e} \tilde F\form{4} \land \tilde F\form{4} \land
  {\mathcal{A}}\form{1} - \frac{1}{e}\, \tilde F_3^{({\textsc{ns}})} \land \tilde
  F\form{3}^{({\textsc{r}})} \land A\form{3}\ .
\end{multline}  
Knowing that both describe the same theory we easily obtain the
conversion table in Tab.~\ref{tab:T-duality} which acts as a
dictionary between IIA and IIB in ten dimensions plus the following
relation between the scalar fields
\begin{equation}
  \begin{pmatrix}
    \Phi \\
    \varPhi 
  \end{pmatrix}_{IIA} =
  \begin{pmatrix}
    3/4  & - \sqrt{7} /4 \\
    -\sqrt{7} / 4 & - 3 / 4
  \end{pmatrix} \begin{pmatrix}
    \Phi \\
    \varPhi 
  \end{pmatrix}_{IIB} 
\end{equation}
This completes the T-duality relations generalizing the usual
ones~\cite{Buscher:1987sk} valid in the \textsc{ns--ns} sector.

\newcolumntype{S}{>{\centering\arraybackslash} m{4em} }

\begin{table}
  \centering
  \begin{tabular}{SSSSS} \toprule
    &\multicolumn{2}{c}{IIA} &\multicolumn{2}{c}{IIB} \\ 
    & $D=10$ & \multicolumn{2}{c}{$D=9$} &  $D=10$ \\ \midrule
    \multirow{4}{4em}{\textsc{r--r} fields}& \multirow{2}{*}{$A_3$} & $A_3$ &     $A_3$ & $B_4$ \\ 
     & &  $A_2^{(2)}$&  $A_2^{\textsc{r}}$ & \multirow{2}{*}{$A_2^{\textsc{r}}$} \\ 
    & \multirow{2}{*}{${\mathcal{A}}_1^{(1)}$} & ${\mathcal{A}}_1^{(1)}$ & $A_1^{\textsc{r}}$ & \\ 
    & & ${\mathcal{A}}_0^{(12)}$ &  $\chi$ &$\chi$  \\ \midrule
    \multirow{4}{4em}{\textsc{ns--ns} fields}& $G_{\mu\nu}$ & ${\mathcal{A}}_1^{(2)}$  & $A_1^{\textsc{ns}}$ & \multirow{2}{*}{$A_2^{\textsc{ns}}$} \\ 
    & \multirow{2}{*}{$A_2^{(1)}$} & $A_2^{(1)}$ &  $A_2^{\textsc{ns}}$ &  \\ 
    & & $A_1^{(12)}$ &  ${\mathcal{A}}_1$ & $G_{\mu\nu}$   \\ \bottomrule
  \end{tabular}
  \caption{T--duality dictionary with \textsc{r--r} fields}
  \label{tab:T-duality}
\end{table}

\section{Killing spinor equations in Type IIB SUGRA}\label{a}
Type IIB supergravity contains a dilatino $\lambda$ and gravitino
$\psi_M$ which can be expressed in terms of complex Weyl spinors. At
tree level, unbroken supersymmetries are manifested by the invariance
of the fermionic fields under the possible supersymmetry
transformations. Supersymmetric configurations satisfy the following
equations
\begin{subequations}
  \label{killingiib}
  \begin{align}
    \delta\lambda &= i\Gamma^M\epsilon^*P_M-\frac{i}{24}\Gamma^{KLN}G_{KLN}\,\,\epsilon=0 \, , \\
    \delta\psi_M &={\cal
      D}_M\epsilon+\frac{1}{96}\left(\Gamma_M{}^{KLN}G_{KLN}-9\Gamma^{LN}G_{MLN}\right)\epsilon^*
    +\frac{i}{4\cdot 480}\Gamma^{M_1\cdots M_5}F_{M_1\cdots
      M_5}\Gamma_M\epsilon =0 \, ,
  \end{align}
\end{subequations}
where we have chosen the complex Weyl spinors to satisfy
$\Gamma^{(11)}\psi_M=\psi_M,$ $\Gamma^{(11)}\epsilon=\epsilon,$ and
$\Gamma^{(11)}\lambda=-\lambda$. The equations above are expressed in
Einstein frame~\cite{Schwarz:1983qr}.

Let us now build the Killing spinors in the $\AdS_3\times S^3\times
T^4$ background. The vielbein for the Einstein metric reads
\begin{align}
\begin{array}{lll}
e^{(0)}=R\cosh(\omega)d\tau &\quad e^{(1)}=R d\omega &\quad e^{(2)}=R\sinh{\omega}d\tau+Rd\sigma \\
e^{(3)}=Rd\theta &\quad e^{(4)}=R\sin(\theta)d\psi &\quad e^{(5)}=R\cos{\theta}d\theta
+Rd\phi \\
e^{(i)}=d x_i &\quad(i=6,\cdots, 8) &\quad e^{(9)}=d\zeta
\end{array}
\end{align}
(Note that here we have rescaled the torus coordinates with respect to
the metric in (\ref{eq:IIB-metric})). The covariant derivatives are
\begin{subequations}
  \begin{align}
    {\cal D}_\tau&=\partial_\tau - \frac{1}{4}\left[
      \sinh{\omega}\gamma^{01}+\cosh{\omega}\gamma^{12}\right] \, , &
    {\cal D}_\omega &= \partial_\omega+\frac{1}{4}\gamma^{02} \, , &
    {\cal D}_\sigma &= \partial_\sigma+\frac{1}{4}\gamma^{01} \, , \\
    {\cal D}_\psi &= \partial_\psi-\frac{1}{4}\left[\cos{\theta}\gamma^{34}-\sin{\theta}\gamma^{35}\right] \, , & {\cal D}_\theta &= \partial_\theta-\frac{1}{4}\gamma^{45} \, , & {\cal D}_\phi &= \partial_\phi+\frac{1}{4}\gamma^{34} \\
    {\cal D}_i &= \partial_i \, , (i=6,\cdots, 9).
  \end{align}
\end{subequations}
where we indicate with lower case greek letters the Dirac matrices in the orthonormal frame (tangent space). The gamma matrices in the coordinate frame $\Gamma_{M}$ can be expressed in terms of the gamma matrices $\gamma_{a}$ in the orthonormal frame as
\begin{subequations}
  \begin{align}
    \Gamma_\tau &= R\left[ \cosh{\omega}\gamma_{0} +
      \sinh{\omega}\gamma_{2} \right] &
    \Gamma_\omega &= R \gamma_{1} &  \Gamma_\sigma &= R\gamma_{2} \\
    \Gamma_\theta &= R\gamma_{3} & \Gamma_\psi &= R\left[\sin{\theta}\gamma_{4}+\cos{\theta}\gamma_5 \right] & \Gamma_\phi &= R\gamma_{5} \\
    \Gamma_i &= \gamma_{i},\quad(i=6,\dots, 9) \, ,
  \end{align}
\end{subequations}
and
\begin{subequations}
  \begin{align}
    \Gamma^\tau &= \frac{1}{R}\sech{\omega}\gamma^{0} & \Gamma^\omega
    &= \frac{1}{R}\gamma^{1} &
    \Gamma^\sigma &= \frac{1}{R}\left[-\tanh{\omega}\gamma^{0}+\gamma^{2}\right],\\
    \Gamma^\theta &= \frac{1}{R}\gamma^{3} & \Gamma^\psi &=
    \frac{1}{R}\csc{\theta}\gamma^{4} &
    \Gamma^\phi &= -\frac{1}{R}\left[\cot{\theta}\gamma^{4}-\gamma^{5}\right] \\
    \Gamma^i &= \gamma^{i},\quad (i = 6,\dots, 9) \, .
  \end{align}
\end{subequations}
The $SU(1,1)$-invariant three-form field $G$ is given by
\begin{equation}
  G\form{3}=\frac{i}{\sqrt{\tau_{2}}}\left( \di C\form{2} - \tau \, \di B\form{2} \right)
\end{equation}
where $\tau = \tau_{1} + i \tau_{2} = C\form{0} + i e^{-\Phi_{IIB}}$ is the
complex scalar field which contains the \textsc{r--r} zero-form $C\form{0}$ and the
dilaton $\Phi_{IIB}$, $C\form{2}$ is the \textsc{r--r} two-form and $B\form{2}$ is the
NS two-form field. For our field configuration:
\begin{equation}
  G\form{3}=iR^2 \left[ \sin{\theta} \di \theta \wedge \di \psi \wedge \di \phi + \cosh{\omega} \di \tau \wedge \di \omega \wedge \di \sigma  \right] \, ,
\end{equation}
and we can set the dilaton to zero. The dilatino variation reads 
\begin{subequations}
    \label{adsdilatino}
  \begin{align}
    \delta\lambda &= -\frac{i}{24}\Gamma^{KLM}G_{KLM}=0  \\
    \left(\Gamma^{\theta\phi\psi}G_{\theta\phi\psi}+\Gamma^{\tau\omega\sigma}G_{\tau\omega\sigma}\right)\epsilon
    &=i \frac{1}{R}\left(\gamma^{354}+\gamma^{012}\right)\epsilon=0
  \end{align}
\end{subequations}
so that
\begin{equation}
  \left(\gamma^{012345}+1\right)\epsilon = 0
\end{equation}
The gravitino conditions are
\begin{align}
\label{adsgravitino}
\begin{split}
  \delta\psi_t&=\partial_t\epsilon-\frac{1}{4}\left[\sinh{\omega}\gamma^{01}+\cosh{\omega}\gamma^{12}\right]\epsilon \\ & \quad + \frac{i}{16}\left[-\cosh{\omega}\left(\gamma^{0345}+3\gamma^{12}\right)+\sinh{\omega}(\gamma^{2345}-3\gamma^{01})\right]\epsilon^*=0,
\end{split}
\nonumber\\
\delta\psi_\omega&=\partial_\omega\epsilon+\frac{1}{4}\gamma^{02}\epsilon+\frac{i}{16}\left[3\gamma^{02}+\gamma^{1345}\right]\epsilon^*=0,\nonumber\\
\delta\psi_\sigma&=\partial_\sigma\epsilon+\frac{1}{4}\gamma^{01}\epsilon+\frac{i}{16}\left[-3\gamma^{01}+\gamma^{2345}\right]\epsilon^*=0,\nonumber\\
\delta\psi_\theta&=\partial_\theta\epsilon-\frac{1}{4}\gamma^{45}\epsilon-\frac{i}{16}\left[3\gamma^{45}+\gamma^{0123}\right]\epsilon^*=0,\nonumber\\
\begin{split}
  \delta\psi_\psi&=\partial_\psi\epsilon+\frac{1}{4}\left[\sin{\theta}\gamma^{35}-\cos{\theta}\gamma^{34}\right]\epsilon \\ & \quad + \frac{i}{16}\left[\sin{\theta}\left(-\gamma^{0124}+3\gamma^{35}\right)-\cos{\theta}(\gamma^{0125}+3\gamma^{34})\right]\epsilon^*=0,
\end{split}
\nonumber\\
\delta\psi_\phi&=\partial_\phi\epsilon+\frac{1}{4}\gamma^{34}\epsilon-\frac{i}{16}\left[3\gamma^{34}+\gamma^{0125}\right]\epsilon^*=0,\nonumber\\
\delta x_i&=\partial_i\epsilon=0, \quad (i=6,\cdots 9).
\end{align}
The last equation tells us that the Killing spinors are constant
around $T^{4}$. Imposing the dilatino condition (\ref{adsdilatino}),
the previous equations simplify to\footnote{Notice that
  $\left(\gamma^{012345}+1\right)\epsilon^{*}=0$}
\begin{equation}
  \partial_\mu
  \begin{pmatrix}
    \epsilon_1\\ \epsilon_2
  \end{pmatrix} =\frac{1}{2} \Omega_\mu
  \begin{pmatrix}
    \epsilon_1\\ \epsilon_2
  \end{pmatrix}
\label{eqkillingIIB}
\end{equation}
with
\begin{subequations}
  \begin{align}
    \Omega_t &= \left(\sinh{\omega}\gamma^{01}+
      \cosh{\omega}\gamma^{12}\right)\otimes \mathcal{P}_{+} &
    \Omega_\omega &= -\gamma^{02}\otimes \mathcal{P}_+ &
    \Omega_\sigma &= -\gamma^{01}\otimes \mathcal{P}_{-} \\
    \Omega_\psi &=
    \left(-\sin{\theta}\gamma^{35}+\cos{\theta}\gamma^{34}\right)\otimes\mathcal{P}_{+} &
    \Omega_\theta &= \gamma^{45}\otimes \mathcal{P}_{+} &
    \Omega_\phi &= -\gamma^{34}\otimes \mathcal{P}_{-}
  \end{align}
\end{subequations}
where $\epsilon^{*}=-i\sigma^{1}\epsilon$ and we have defined the
projectors $\mathcal{P}_{\pm}=\frac{\mathbb{1}\pm\sigma^{1}}{2}$. Note
that $\Omega_{\mu}=A_{(32)}\otimes \mathcal{P}_{\pm}$ so that $A_{32}$
is a $32\times32$ matrix which has the property $A^{2}=\pm 1$ for
fixed value of $\mu$. Hence schematically $\Omega_{\mu}^{2}=\pm
\mathbb{1}_{32} \otimes \mathcal{P}$. Making use of these facts and
the following identity
\begin{equation}
  \exp{\Omega x}=(\mathbb{1}_{32}\otimes \bar{\mathcal{P}}) + (\cosh{x}+A \sinh{x})\otimes \mathcal{P}
\end{equation}
where $\bar{\mathcal{P}}=\mathbb{1}-\mathcal{P}$, one can integrate
(\ref{eqkillingIIB}) and solve for the Killing spinor $\epsilon$. We
write
$\epsilon=\epsilon_{\omega}\epsilon_{t}\epsilon_{\sigma}\epsilon_{\theta}\epsilon_{\psi}\epsilon_{\phi}$. The
Killing spinor has the form:
\begin{equation}
  \epsilon=e^{-\frac{\omega}{2}\gamma^{02}\otimes\mathcal{P_{+}}}e^{\frac{t}{2}\gamma^{12}\otimes\mathcal{P_{+}}}e^{\frac{\sigma}{2}\gamma^{01}\otimes\mathcal{P_{-}}}
  e^{\frac{\theta}{2}\gamma^{45}\otimes\mathcal{P_{+}}}e^{\frac{\psi}{2}\gamma^{34}\otimes\mathcal{P_{+}}}e^{\frac{-\phi}{2}\gamma^{34}\otimes\mathcal{P_{-}}}\epsilon_0
\end{equation}
The constant spinors $\epsilon_0$ arising from the integration
constants, is constrained by the projection
\begin{align}
\left(\gamma^{012345}+1\right)\epsilon_0=0.
\end{align}
so that 16 supersymmetries remain unbroken\footnote{This is just the
  usual $T^4$ projection, $\Gamma^{6789}\epsilon_0=\epsilon_0$.}.
\subsection{Squashed Sphere and Warped AdS}
The field configuration for this type IIB background that arises after
two Hopf T-dualities is determined by eqs. (\ref{twotdualities}). The
covariant derivatives are:
\begin{align}
  \mathcal{D}_{\tau}&=\partial_{\tau}+\frac{(\sech^2{\Theta_{w}}-2)}{4}\sinh{\omega}\gamma^{01}-\frac{\sech{\Theta_{w}}}{4}\cosh{\omega}\gamma^{12} \qquad \qquad \mathcal{D}_{\omega}=\partial_{\omega}+\frac{\sech\Theta_{w}}{4}\gamma^{02} \nonumber \\
  \mathcal{D}_{\beta}&=\partial_{\beta}+\frac{\sech^2{\Theta_{w}}}{4}\gamma^{01} \qquad \mathcal{D}_{\theta}=\partial_{\theta}-\frac{\sech\Theta_m}{4}\gamma^{45} \qquad \mathcal{D}_{i}=\partial_{i}, \qquad i=\zeta_m, \zeta_w, 8, 9 \nonumber \\
  \mathcal{D}_{\phi}&=\partial_{\phi}+\frac{(\sech^2\Theta_{m}-2)}{4}\cos{\theta}\gamma^{34}+\frac{\sech\Theta_{m}}{4}\sin{\theta}\gamma^{35} \qquad \qquad \mathcal{D}_{\alpha}=\partial_{\alpha}
+\frac{\sech^2\Theta_{m}}{4}\gamma^{34}
\end{align}
and the gamma matrices in the coordinate frame $\Gamma_{M}$ read
\begin{align}
\Gamma_{\tau}&=R(-\cosh{\omega}\gamma^{0}+\sech\Theta_{w} \sinh{\omega}\gamma^{2}) \qquad \qquad \Gamma_{\phi}=R(\sin{\theta}\gamma^{4}+\sech\Theta_{m} \cos{\theta}\gamma^{5})
\nonumber \\
\Gamma_{\omega}&=R \gamma^{1}\qquad \qquad \Gamma_{\theta}=R \gamma^{3} \qquad \qquad \Gamma_{\beta}=R \sech\Theta_{w}\gamma^{2} \qquad \Gamma_{\alpha}=\sech\Theta_{m}\gamma^{5} \nonumber \\
\Gamma_{\zeta_m}&=\gamma^6 \qquad \qquad \Gamma_{\zeta_w}=\gamma^{7} \qquad \qquad 
 \Gamma_{i}=\gamma^{i} \qquad i=8,9
\end{align}
and
\begin{align}
\Gamma^{\tau}&=\frac{1}{R}\sech{\omega}\gamma^{0} \qquad \Gamma^{\phi}=\frac{1}{R}\csc{\theta}\gamma^{4}
\qquad
\Gamma^{\omega}=\frac{1}{R}\gamma^{1} \hspace{18.mm} \Gamma^{\theta}=\frac{1}{R}\gamma^{3} \nonumber \\
\Gamma^{\beta}&=\frac{1}{R}(-\tanh{\omega}\gamma^0+\cosh{\Theta_{w}}\gamma^2) \qquad \Gamma^{\alpha}=\frac{1}{R}(-\cot{\theta}\gamma^4+\cosh{\Theta_m}\gamma^5) \nonumber \\
\Gamma^{\zeta_m}&=\gamma^6 \qquad \qquad \Gamma^{\zeta_w}=\gamma^{7} \qquad \qquad 
 \Gamma^{i}=\gamma^{i} \qquad i=8,9
\end{align}
The three-form field $G_{3}$ reads in this case
\begin{equation}
  G_{3}=-R \tanh{\Theta_{w}}\cosh{\omega}d\tau \wedge d \omega \wedge (i d\zeta_m+d\zeta_w)+R \tanh{\Theta_{m}}\sin{\theta}d\theta \wedge d \phi \wedge (id \zeta_w+d\zeta_m)
\end{equation}

\section{Killing spinor equations in Type IIA SUGRA}\label{iia}
The supersymmetry variations for the dilatino and the gravitino for type IIA supergravity read
\begin{align}
  \begin{split}
    \sqrt{2}\delta\lambda&=-\frac{1}{2}{\cal D}_M\Phi_{IIA}\,\Gamma^M\Gamma^{11}\epsilon+\frac{1}{24}e^{-\frac{\Phi_{IIA}}{2}}H^{(3)}_{M_1M_2M_3}\Gamma^{M_1M_2M_3}\epsilon \\
    & \quad -\frac{3}{8\cdot
      2!}e^{\frac{3\Phi_{IIA}}{4}}F^{(2)}_{M_1M_2}\Gamma^{M_1M_2}\epsilon+\frac{1}{8\cdot
      4!}e^{\frac{\Phi_{IIA}}{4}}F^{(4)}_{M_1M_2M_3M_4}\Gamma^{M_1M_2M_3M_4}\Gamma^{11}\epsilon=0 \,
  \end{split} \\
  \begin{split}
    \delta\psi_M&={\cal D}_M\epsilon+\frac{1}{96}e^{-\frac{\Phi_{IIA}}{2}}H^{(3)}_{M_1M_2M_3}\left(\Gamma_M{}^{M_1M_2M_3}-9\delta_M^{M_1}\Gamma^{M_2M_3}\right)\Gamma^{11}\epsilon \\
    & \quad -\frac{1}{64}e^{\frac{3\Phi_{IIA}}{4}}F^{(2)}_{M_1M_2}\left(\Gamma_M{}^{M_1M_2}-14\delta_M^{M_1}\Gamma^{M_2}\right)\Gamma^{11}\epsilon \\
    & \quad +\frac{1}{256}e^{\frac{\Phi_{IIA}}{4}}F^{(4)}_{M_1M_2M_3M_4}\left(\Gamma_M{}^{M_1M_2M_3M_4}-\frac{20}{3}\delta_M^{M_1}\Gamma^{M_2M_3M_4}\right)\epsilon=0,
  \end{split}
\end{align}
where the spinor $\epsilon$ is taken to be Majorana but not Weyl. Here $\Phi_{IIA}$ is the dilaton field of type IIA supergravity and the three-form field, $H_{(3)}=dB_{(2)}$ is the NS field strength.

We will consider two cases depending on how the circle chosen to perform T--duality. We first look at the case in which the sphere is squashed, followed by that in which the anti-de Sitter space becomes warped. 
\subsection{Squashed Sphere}
Let the coordinates along the squashed sphere $\SqS^{3}$ be denoted by
$x=\{\theta, \phi, \alpha\}$ (the coordinates on AdS are
$\{\tau,\omega,\sigma\}$). The covariant derivatives along these
directions are given by
\begin{subequations}
  \begin{align}
    \mathcal{D}_{\phi} &= \partial_{\phi}+\frac{(\sech^2\Theta_m-2)}{4}\cos{\theta}\gamma^{34}+\frac{\sech\Theta_m}{4}\sin{\theta} \gamma^{35}  \, , \\
    \mathcal{D}_{\theta}&=\partial_{\theta}-\frac{\sech\Theta_m}{4}\gamma^{45}
    \, , \quad
    \mathcal{D}_{\alpha}=\partial_{\alpha}+\frac{\sech^2\Theta_m}{4}\gamma^{34}
  \end{align}
\end{subequations}
The antisymmetric fields on this basis read
\begin{align}
H_{3}&=R \tanh{\Theta_m}\sin{\theta} d\theta \wedge d\phi \wedge d\zeta_m \\
F_{2}&=R \tanh{\Theta_m}\sin{\theta} d\theta \wedge d\phi  \\
F_{4}&=\left[\frac{R^2}{\cosh{\Theta_m}^2}\sin{\theta} d\theta \wedge d\phi \wedge d\alpha +R^2\cosh{\omega}dt \wedge d\omega \wedge d\sigma \right]\wedge d\zeta_m
\end{align}

The $\Gamma$-matrices in the background are related to those in the orthonormal frame as:
\begin{align}
  \Gamma_{\theta}&=R\gamma^{3} & \Gamma_{\alpha} &= R \sech\Theta_m\gamma^5 &
  \Gamma_{\phi} &= R\left[\sin{\theta}\gamma^4+\sech\Theta_m \cos{\theta}\gamma^5\right] \\
  \intertext{and}
  \Gamma^{\theta} &=\frac{1}{R}\gamma^3 & \Gamma^{\phi} &= \frac{1}{R}\csc{\theta}\gamma^4 &
  \Gamma^{\alpha}&=\frac{1}{R}\left[-\cot{\theta}\gamma^4+\cosh{\Theta_m}\gamma^5\right] \, .
\end{align}
The variations of the gravitino, $\delta \psi_{M}=0$, give raise to the equations:
\begin{small}
  \begin{align}
    \delta \psi_{\phi}&=\left[\partial_{\phi}+\frac{(\sech^2\Theta_m-2)}{4}\cos{\theta}\gamma^{34}+\frac{\sech\Theta_m}{4}\sin{\theta}\gamma^{35}\right]\epsilon \nonumber \\
    &+\frac{1}{32}\tanh{\Theta_m}\left(  \sin{\theta}(6\gamma^{39}-7\gamma^{3})+\sech\Theta_m\cos{\theta}(2\gamma^{3459}-\gamma^{345}) \right)\Gamma^{11}\epsilon \nonumber  \\
    &+\frac{1}{32}\left( \sin{\theta}(5\sech\Theta_m\gamma^{359}-3\gamma^{01249})-\sech\Theta_m\cos{\theta}(5\sech\Theta_m\gamma^{349}+3\gamma^{01259}) \right)\epsilon=0 \nonumber  \\
    \delta\psi_{\alpha}&=\left[\partial_{\alpha}+\frac{\sech^2\Theta_m}{4}\gamma^{34}\right]\epsilon+\frac{1}{32}\tanh{\Theta_m}\left(2\sech\Theta_m\gamma^{3459}-\gamma^{345}\right)\Gamma^{11}\epsilon \nonumber \\
    &-\frac{1}{32}\left(5\sech^2\Theta_m\gamma^{349}+3\sech\Theta_m\gamma^{01259}\right)\epsilon=0 \nonumber  \\
    \delta\psi_{\theta}&=\left[\partial_{\theta}-\frac{\sech\Theta_m}{4}\gamma^{45}\right]\epsilon+\frac{1}{32}\tanh{\Theta_m}\left((-6\gamma^{49}+7\gamma^{4})\right)\Gamma^{11}\epsilon-\frac{1}{32}\left(5\sech\Theta_m\gamma^{459}+3\gamma^{01239}\right)\epsilon=0
    \nonumber \\
    \delta\psi_{\tau}&=\left[\partial_{\tau}-\frac{1}{4}(\sinh{\omega}\gamma^{01}+\cosh{\omega}\gamma^{12})\right]\epsilon \nonumber \\
    &+\frac{1}{32}\tanh{\Theta_m}\left( \sinh{\omega}(2\gamma^{2349}-\gamma^{234})-\cosh{\omega}(2\gamma^{0349}-\gamma^{034})\right)\Gamma^{11}\epsilon \nonumber \\
    &+\frac{1}{32}\left( \sinh{\omega}(3\sech\Theta_m\gamma^{23459}-5\gamma^{019})-\cosh{\omega}(3\sech\Theta_m\gamma^{03459}+5\gamma^{129}) \right)\epsilon=0 \nonumber   \\
    \delta\psi_{\sigma}&=\left[\partial_{\sigma}+\frac{1}{4}\gamma^{01}\right]\epsilon+\frac{1}{32}\tanh{\Theta_m}\left(2\gamma^{2349}-\gamma^{234}\right)\Gamma^{11}\epsilon+\frac{1}{32}\left(3\sech\Theta_m\gamma^{23459}-5\gamma^{019}\right)\epsilon=0 \nonumber \\
    \delta\psi_{\omega}&=\left[\partial_{\omega}+\frac{1}{4}\gamma^{02}\right]\epsilon+\frac{1}{32}\tanh{\Theta_m}\left(2\gamma^{1349}-\gamma^{134}\right)\Gamma^{11}\epsilon+\frac{1}{32}\left(3\sech\Theta_m\gamma^{13459}+5\gamma^{029}\right)\epsilon =0 \nonumber  \\
    \delta\psi_{i}&=\partial_{i}\epsilon+\frac{1}{32}\frac{\tanh{\Theta_m}}{R}\left(2\gamma^{i349}-\gamma^{i34}\right)\Gamma^{11}\epsilon+\frac{3}{32}\frac{1}{R}\left(\sech\Theta_m\gamma^{i3459}+\gamma^{i0129}\right)\epsilon=0 \nonumber  \\
    \delta\psi_{\zeta_w}&=\partial_{\zeta_w}\epsilon+\frac{1}{32}\frac{\tanh{\Theta_m}}{R}\left(-6\gamma^{34}-\gamma^{349}\right)\Gamma^{11}\epsilon+\frac{5}{32}\frac{1}{R}\left(\sech\Theta_m\gamma^{345}+\gamma^{012}\right)\epsilon=0
    \label{vargravisquashed}
  \end{align}
\end{small}
and the dilatino variation gives:
\begin{equation}
\tanh{\Theta_m}\left(2\gamma^{349}-3\gamma^{34}\right)\epsilon+\left(\sech\Theta_m\gamma^{3459}+\gamma^{0129}\right)\Gamma^{11}\epsilon=0
\end{equation}
Combining this equation with eqs. (\ref{vargravisquashed}) one obtains:
\begin{align}
\partial_{i}\epsilon&=\frac{1}{2R}\tanh{\Theta_m}\gamma^{34}\mathcal{Q}_{-}\gamma^{i}\Gamma^{11}\epsilon \nonumber \\
\partial_{\zeta_w}\epsilon&=\frac{1}{R}\tanh{\Theta_m}\gamma^{34}\mathcal{Q}_{+}\Gamma^{11}\epsilon \nonumber  \\
\partial_{\sigma}\epsilon&=-\frac{1}{2}\left[\gamma^{01}\mathcal{Q}_{-}+\tanh{\Theta_m}\gamma^{234}\mathcal{Q}_{+}\Gamma^{11}\right] \epsilon \nonumber
 \\
\partial_{\omega}\epsilon&=-\frac{1}{2}\left[\gamma^{02}\mathcal{Q}_{+}+\tanh{\Theta_m}\gamma^{134}\mathcal{Q}_{+}\Gamma^{11}\right] \epsilon \nonumber 
 \\
\partial_{\tau}\epsilon&=\frac{1}{2}\left[(\sinh{\omega}\gamma^{01}+\cosh{\omega}\gamma^{12})\mathcal{Q}_{+}-\tanh{\Theta_m}(\sinh{\omega}\gamma^{234}-\cosh{\omega}\gamma^{034})\mathcal{Q}_{+}\Gamma^{11}\right] \epsilon \nonumber  
\\
\partial_{\theta}\epsilon&=\frac{1}{2}\left[\gamma^{0123}\mathcal{Q}_{+}-\tanh{\Theta_m}\gamma^{4}\mathcal{Q}_{+}\Gamma^{11}\right]\epsilon \nonumber \\
\partial_{\alpha}\epsilon&=-\frac{1}{2}\left[\sech^2\Theta_m\gamma^{34}\mathcal{Q}_{-}+\sech\Theta_m\tanh{\Theta_m}\gamma^{345}\mathcal{Q}_{+}\Gamma^{11}\right]\epsilon \nonumber  \\
\partial_{\phi}\epsilon&=\frac{1}{2}\left[(\sin{\theta}\gamma^{0124}\mathcal{Q}_{+}-(\sech^2\Theta_m\mathcal{Q}_{-}-1)\cos{\theta}\gamma^{34})\right. \nonumber \\
&\left.+\tanh{\Theta_m}(\sin{\theta}\gamma^{3}\mathcal{Q}_{+}-\sech{\Theta_m}\cos{\theta}\gamma^{345}\mathcal{Q}_{+})\Gamma^{11}\right]\epsilon
\end{align}
Here $\mathcal{Q}_{\pm}=\frac{\mathbb{1}\pm \gamma^9}{2}$. These equations all have solutions of the form $\epsilon=\exp{x}\epsilon_0$. One needs to be careful with the boundary conditions when compactifying. Therefore, from the equation for the $S^1$, denoted by the coordinate $z$, one obtains the following constraint on the Killing spinors
\begin{equation}
\gamma^{34}\mathcal{Q}_{+}\Gamma^{11}\epsilon=0 
\end{equation}
Hence the Killing spinors do not depend on the torus coordinates as it should be. Let us write $\epsilon=\epsilon_{1}(\omega,\tau,\sigma)\epsilon_{2}(\theta,\phi,\alpha)\epsilon_0$, with $\epsilon_0$ a constant Majorana spinor. The set of equations for the AdS coordinates becomes
\begin{align}
\partial_{\omega}\epsilon_1&=-\frac{1}{2}\gamma^{02}\mathcal{Q}_{+}\epsilon_1 \\
\partial_{\sigma}\epsilon_1&=-\frac{1}{2}\gamma^{01}\mathcal{Q}_{-}\epsilon_1 \\
\partial_{\tau}\epsilon_1&=\frac{1}{2}\left(\sinh{\omega}\gamma^{01}+\cosh{\omega}\gamma^{12}\right)\mathcal{Q}_{+}\epsilon_1 
\end{align}
Which can be immediately solved as we did on the previous section, by writing $\epsilon_1(\omega,\tau,\sigma)=\epsilon_{1\omega}(\omega)\epsilon_{1\tau}(\omega,\tau)\epsilon_{1\sigma}(\omega,\tau,\sigma)$ (they are just the Killing spinor equations for $AdS_{3}$). One obtains
\begin{align}
\epsilon_{1}(\omega,\tau,\sigma)=\exp\left\{-\frac{\omega}{2}\gamma^{02}\mathcal{Q}_{+}\right\}\exp\left\{\frac{\tau}{2}\gamma^{12}\mathcal{Q}_{+}\right\}\exp\left\{\frac{\sigma}{2}\gamma^{01}\mathcal{Q}_{-}\right\}
\end{align}
The remaining equations can be solved in analogous fashion. Let $\epsilon_{2}(\theta,\phi,\alpha) = \epsilon_{2\theta}(\theta) \epsilon_{2\phi}(\theta,\phi) \epsilon_{2\alpha}(\theta,\psi,\alpha)$. The solution is:
\begin{multline}
  \epsilon_{2}(\theta,\phi,\alpha) = \exp\left\{\frac{\theta}{2}\gamma^{0123}\mathcal{Q}_{+}\right\}\left[\exp\left\{ \tanh^2{\Theta_m}\cos{\theta}\left(\frac{\phi}{2}\right)\gamma^{34} \right\}\mathcal{Q}_{-} \right.  \\
  +\left.\exp\left\{\frac{\phi}{2}\gamma^{34}
    \right\}\mathcal{Q}_{+}\right]\exp\left\{-\sech^2\Theta_{m}\frac{\alpha}{2}\gamma^{34}\mathcal{Q}_{-}\right\}
\end{multline}
If one writes $\epsilon=\epsilon_{+}+\epsilon_{-}$ with $\mathcal{Q}_{\pm}\epsilon_{\pm}=\pm\epsilon_{\pm}$  one can verify that only the supersymmetries associated to the $\epsilon_{-}$ spinors are preserved after T-duality. Hence the solution is $1/4$-BPS. When the deformation parameter is such that $\sech{\Theta_m}=0$, there are no spinors depending on the T-dual coordinate and supersymmetry is restored to $1/2$-BPS. This corresponds to the case in which the geometry becomes $AdS_{3}\times S^{2}\times S^1\times T^4$
\subsection{Warped AdS}
Let the coordinates along the warped anti-de Sitter space $WAdS_{3}$
be denoted by $x=\{t, \omega, \alpha\}$ (the coordinates on the sphere
read $\{\theta,\psi,\phi \}$). The covariant derivatives along these
directions are given by
\begin{subequations}
  \begin{align}
    \mathcal{D}_{\tau} &= \partial_{\tau}+\frac{(\sech^2\Theta_w-2)}{4}\sinh{\omega}\gamma^{01}-\frac{\sech\Theta_w}{4}\cosh{\omega}\gamma^{12} \, , \\
    \mathcal{D}_{\omega}
    &= \partial_{\omega}+\frac{\sech\Theta_w}{4}\gamma^{02} \, , \quad
    \mathcal{D}_{\beta}=\partial_{\beta}+\frac{\sech^2\Theta_w}{4}\gamma^{01}
    \, .
  \end{align}
\end{subequations}
The antisymmetric fields on this basis read
\begin{align}
H_{3}&=-R \tanh{\Theta_w}\cosh{\omega} d\tau \wedge d\omega \wedge d\zeta_w \\
F_{2}&=-R \tanh{\Theta_w}\cosh{\omega} d\tau \wedge d\omega  \\
F_{4}&=\left[-R^2\sech^2\Theta_w\cosh{\omega} d\tau \wedge d\omega \wedge d\beta + R^2\sin{\theta}d\theta \wedge d\psi \wedge d\phi \right] \wedge d\zeta_w
\end{align}
The $\Gamma$-matrices in the background are related to those in the orthonormal frame as:
\begin{align}
  \Gamma_{\omega} &=R\gamma^{1} & \Gamma_{\beta} &= R\sech\Theta_w\gamma^2 &
  \Gamma_{\tau} &= R\left[-\cosh{\omega}\gamma^0+\sech\Theta_w \sinh{\omega}\gamma^2\right] \\
  \intertext{and}
  \Gamma^{\omega} &= \frac{1}{R}\gamma^1 & \Gamma^{\tau} &= \frac{1}{R}\sech{\omega}\gamma^0 &
  \Gamma^{\beta} &= \frac{1}{R}\left[-\tanh{\omega}\gamma^0+\cosh{\Theta_w}\gamma^2\right] \, .
\end{align}
The variations of the gravitino, $\delta \psi_{M}=0$, give rise to the equations:
\begin{small}
  \begin{align}
    \begin{split}
      \delta \psi_{\tau}&=\left[\partial_{\tau}+\frac{(\sech^2\Theta_w-2)}{4}\sinh{\omega}\gamma^{01}-\frac{\sech\Theta_w}{4}\cosh{\omega}\gamma^{12}\right]\epsilon  \\
      & \quad +\frac{1}{32}\tanh{\Theta_w}\left(\cosh{\omega}(6\gamma^{19}-7\gamma^{1})+\sech\Theta_w\sinh{\omega}(-2\gamma^{0129}+\gamma^{012})\right)\Gamma^{11}\epsilon  \\
      & \quad +\frac{1}{32}\left(\cosh{\omega}(5\sech\Theta_w\gamma^{129}-3\gamma^{03459})+\sech\Theta_w\sinh{\omega}(5\sech\Theta_w\gamma^{019}+3\gamma^{23459})\right)\epsilon=0
    \end{split}
\nonumber  \\
    \delta\psi_{\omega}&=\left[\partial_{\omega}+\frac{\sech\Theta_w}{4}\gamma^{02}\right]\epsilon+\frac{1}{32}\tanh{\Theta_w}\left(-6\gamma^{09}+7\gamma^{0}\right)\Gamma^{11}\epsilon
    +\frac{1}{32}\left(-5\sech\Theta_w\gamma^{029}+3\gamma^{13459}\right)\epsilon=0 \nonumber  \\
    \begin{split}
      \delta\psi_{\beta}&=\left[ \partial_{\beta}+\frac{\sech^2\Theta_w}{4}\gamma^{01}\right]\epsilon+\frac{1}{32}\tanh{\Theta_w}\left(-2\sech\Theta_w\gamma^{0129}+\gamma^{012}\right)\Gamma^{11}\epsilon
       \\
      & \quad +\frac{1}{32}\left(5\sech^2\Theta_w \gamma^{019}+3\sech\Theta_w
        \gamma^{23459}\right)\epsilon=0
    \end{split}
\nonumber \\
    \delta\psi_{\theta}&=\left[ \partial_{\theta}-\frac{1}{4}\gamma^{45}\right]\epsilon+\frac{1}{32}\tanh{\Theta_w}\left(-2\gamma^{0139}+\gamma^{013}\right)\Gamma^{11}\epsilon+\frac{1}{32}\left(3\sech\Theta_w\gamma^{01239}-5\gamma^{459}\right)\epsilon=0 \nonumber  \\
    \begin{split}
      \delta\psi_{\psi}&=\left[ \partial_{\psi}-\frac{1}{4}(\cos{\theta}\gamma^{34}+\sin{\theta}\gamma^{35})\right]\epsilon \\
      & \quad +\frac{1}{32}\tanh{\Theta_w}\left(\sin{\theta}(-2\gamma^{0149}+\gamma^{014})+\cos{\theta}(-2\gamma^{0159}+\gamma^{015})\right)\Gamma^{11}\epsilon \nonumber \\
      & \quad +\frac{1}{32}\left(\sin{\theta}(3\sech\Theta_w\gamma^{01249}+5\gamma^{359})+\cos{\theta}(3\sech\Theta_w\gamma^{01259}-5\gamma^{349})\right)\epsilon=0
    \end{split}
\nonumber  \\
    \delta\psi_{\phi}&=\left[\partial_{\phi}+\frac{1}{4}\gamma^{34}\right]\epsilon+\frac{1}{32}\tanh{\Theta_w}\left(-2\gamma^{0159}+\gamma^{015}\right)\Gamma^{11}\epsilon+\frac{1}{32}\left(3\sech\Theta_w\gamma^{01259}-5\gamma^{349}\right)\epsilon=0 \nonumber \\
    \delta\psi_{i}&=\partial_{i}\epsilon+\frac{1}{32}\frac{1}{R}\tanh{\Theta_w}\left(-2\gamma^{01i9}+\gamma^{01i}\right)\Gamma^{11}\epsilon+\frac{3}{32R}\left(\sech\Theta_w\gamma^{012i9}-\gamma^{345i9}\right)\epsilon=0 \nonumber \\
    \delta\psi_{\zeta_w}&=\partial_{\zeta_w}\epsilon+\frac{1}{32}\frac{1}{R}\tanh{\Theta_w}\left(6\gamma^{01}+\gamma^{019}\right)\Gamma^{11}\epsilon+\frac{5}{32R}\left(-\sech\Theta_w\gamma^{012}+\gamma^{345}\right)\epsilon=0
  \end{align}
\end{small}
and the variation of the dilatino
\begin{equation}
  \tanh{\Theta_w}(-2\gamma^{019}+3\gamma^{01})\epsilon+(-\sech\Theta_w\gamma^{0129}+\gamma^{3459})\Gamma^{11}\epsilon=0
  \label{dilatinowads}
\end{equation}
Using (\ref{dilatinowads}) into the previous equations, yields
\begin{align}
\partial_{i}\epsilon&=\frac{1}{2R}\tanh{\Theta_w}\gamma^{01i}\mathcal{Q}_{+}\Gamma^{11}\epsilon \nonumber \\
\partial_{\zeta_w}\epsilon&=-\frac{1}{R}\tanh{\Theta_w}\gamma^{01}\mathcal{Q}_{+}\Gamma^{11}\epsilon \nonumber  \\
\partial_{\theta}\epsilon&=\frac{1}{2}\left[\gamma^{45}\mathcal{Q}_{+}+\tanh{\Theta_w}\gamma^{013}\mathcal{Q}_{+}\Gamma^{11}\right]\epsilon \nonumber  \\
\partial_{\psi}\epsilon&=\frac{1}{2}\left[(-\sin{\theta}\gamma^{35}+\cos{\theta}\gamma^{34})\mathcal{Q}_{+}+\tanh{\Theta_w}(\sin{\theta}\gamma^{014}+\cos{\theta}\gamma^{015})\mathcal{Q}_{+}\Gamma^{11}\right]\epsilon \nonumber  \\
\partial_{\phi}\epsilon&=-\frac{1}{2}\left[\gamma^{34}\mathcal{Q}_{-}-\tanh{\Theta_w}\gamma^{015}\mathcal{Q}_{+}\Gamma^{11}\right]\epsilon \nonumber  \\
\partial_{\beta}\epsilon&=-\frac{1}{2}\left[ \sech\Theta_w^2\gamma^{01}\mathcal{Q}_{+}-\sech\Theta_w\tanh{\Theta_w}\gamma^{012}\mathcal{Q}_{+}\Gamma^{11}\right]\epsilon \nonumber  \\
\partial_{\omega}\epsilon&=-\frac{1}{2}\left[ -\gamma^{1345}\mathcal{Q}_{-}+\tanh{\Theta_w}\gamma^{0}\mathcal{Q}_{+}\Gamma^{11} \right]\epsilon \nonumber  \\
\begin{split}
  \partial_{\tau}\epsilon&=\frac{1}{2}\left[
    (-\cosh{\omega}\gamma^{0345}\mathcal{Q}_{-}-(\sech^2\Theta_w\mathcal{Q}_{+}-1)\sinh{\omega}\gamma^{01})\right.
   \\
  & \quad \left.+\tanh{\Theta_w}(\sech\Theta_w \sinh{\omega}
    \gamma^{012}\mathcal{Q}_{+}-\cosh{\omega}\gamma^{1}\mathcal{Q}_{+})\Gamma^{11}\right]\epsilon
\end{split}
\end{align}
Again, these equations have solutions of the form $\epsilon=\exp{x}\epsilon_0$. As before, from the equation for the $z$ coordinate, one obtains the constraint on the Killing spinors
\begin{equation}
\gamma^{01}\mathcal{Q}_{+}\Gamma^{11}\epsilon=0 
\end{equation}
We write $\epsilon=\epsilon_{1}(\theta,\psi,\phi)\epsilon_{2}(\omega,\tau,\beta)\epsilon_0$, with $\epsilon_0$ a constant Majorana spinor. The set of equations for the sphere coordinates becomes
\begin{align}
\partial_{\theta}\epsilon_1&=\frac{1}{2}\gamma^{45}\mathcal{Q}_{+}\epsilon_1 \\
\partial_{\phi}\epsilon_1&=-\frac{1}{2}\gamma^{34}\mathcal{Q}_{-}\epsilon_1 \\
\partial_{\psi}\epsilon_1&=\frac{1}{2}\left(-\sin{\theta}\gamma^{35}+\cos{\theta}\gamma^{34}\right)\mathcal{Q}_{+}\epsilon_1 
\end{align}
Which can be immediately solved as we did on the previous section, by writing $\epsilon_1(\theta,\psi,\phi)=\epsilon_{1\theta}(\theta)\epsilon_{1\psi}(\theta,\psi)\epsilon_{1\phi}(\theta,\psi,\phi)$ (they are just the Killing spinor equations for $S^{3}$). One obtains
\begin{align}
\epsilon_{1}(\theta,\psi,\phi)=\exp\left\{\frac{\theta}{2}\gamma^{45}\mathcal{Q}_{+}\right\}\exp\left\{\frac{\psi}{2}\gamma^{34}\mathcal{Q}_{+}\right\}\exp\left\{-\frac{\phi}{2}\gamma^{34}\mathcal{Q}_{-}\right\}
\end{align}
The remaining equations can be solved in analogous fashion. Let 
\begin{equation}
  \epsilon_{2}(\omega,\tau,\beta) = \epsilon_{2\omega}(\omega) \epsilon_{2\tau}(\omega,\tau) \epsilon_{2\beta}(\omega,\tau,\beta) \, .    
\end{equation}
The solution is:
\begin{multline}
  \epsilon_{2}(\omega,\tau,\beta) = \exp\left\{-\frac{\omega}{2}\gamma^{1345}\mathcal{Q}_{-}\right\}\left[\exp\left\{ -\tanh{\Theta_w}^2\sinh{\omega}\left(\frac{\tau}{2}\right)\gamma^{01} \right\}\mathcal{Q}_{+} \right. \\
   +\left.\exp\left\{\frac{\tau}{2}\gamma^{0345} \right\}\mathcal{Q}_{-}\right]\exp\left\{-\sech^2\Theta_w\frac{\beta}{2}\gamma^{01}\mathcal{Q}_{+}\right\}
\end{multline}
The arguments given in the previous section also apply to this
case. For $\epsilon=\epsilon_{+}+\epsilon_{-}$ with
$\mathcal{Q}_{\pm}\epsilon_{\pm}=\pm\epsilon_{\pm}$ only the
supersymmetries associated to the $\epsilon_{+}$ spinors are preserved
after T-duality and the solution is $1/4$-BPS. When the deformation
parameter is such that $\sech{\Theta_w}=0$, there are no spinors
depending on the T-dual coordinate and supersymmetry is restored to
$1/2$-BPS. This corresponds to the case in which the geometry becomes
$\AdS_{2}\times S^{1}\times S^3\times T^4$
\end{document}